\documentclass[epj]{svjour}
\usepackage[]{fontenc}
\usepackage[latin1]{inputenc}
\usepackage{amsmath}
\usepackage{color}
\usepackage{graphicx}
\usepackage{amssymb}

\makeatletter

\providecommand{\tabularnewline}{\\}

\usepackage{epsfig}
\newcommand{\be}{\begin{equation}}
\newcommand{\bea}{\begin{eqnarray}}
\newcommand{\eea}{\end{eqnarray}}
\newcommand{\ee}{\end{equation}}

\makeatother
\begin{document}

\title{Stable particles in anisotropic spin-1 chains}

\author{L. Campos Venuti\inst{1,3} \and C. Degli Esposti Boschi\inst{1,4} \and%
E. Ercolessi\inst{2,3,5} \and G. Morandi\inst{1,3,5} \and
F. Ortolani\inst{2,3,5} \and S. Pasini\inst{2,3} \and M. Roncaglia\inst{1,3} 
}
\mail{cristian.degliesposti@unibo.it}

\institute{Dipartimento di Fisica dell'Università di Bologna, viale
Berti-Pichat, 6/2, 40127, Bologna, Italia \and 
Dipartimento di Fisica dell'Università di Bologna, via Irnerio 46,
40126, Bologna, Italia \and
INFN, Sezione di Bologna, viale Berti-Pichat, 6/2, 40127, Bologna,
Italia \and
CNR-INFM, Unità di Ricerca di Bologna, viale Berti-Pichat, 6/2, 40127,
Bologna, Italia \and
CNISM, Sezione di Bologna, viale Berti-Pichat, 6/2, 40127,
Bologna, Italia
}

\abstract{
Motivated by field-theoretic predictions we investigate the stable
excitations that exist in two characteristic gapped phases of a spin-1
model with Ising-like and single-ion anisotropies. The sine-Gordon
theory indicates a region close to the phase boundary where a stable
breather exists besides the stable particles, that form the Haldane
triplet at the Heisenberg isotropic point. The numerical data, obtained
by means of the Density Matrix Renormalization Group, confirm this
picture in the so-called large-$D$ phase for which we give also a
quantitative analysis of the bound states using standard perturbation
theory. However, the situation turns out to be considerably more intricate
in the Haldane phase where, to the best of our data, we do not observe
stable breathers contrarily to what could be expected from the sine-Gordon
model, but rather only the three modes predicted by a novel anisotropic
extension of the Non-Linear Sigma Model studied here by means of a
saddle-point approximation. 
} 

\maketitle

\section{Introduction\label{sec:Intro}}

One-dimensional spin systems have been largely studied since Haldane
\cite{haldane83} proposed his conjecture. He suggested that while
half-odd integer spin chains should always be gapless, integer-spin
ones should manifest a gap. The conjecture is supported by a field-theoretic
analysis \cite{aue} of the lattice Hamiltonians, a typical approach
for low-di\-men\-sion\-al quantum systems that admit a continuum-limit
counterpart 
in $(D+1)$ dimensions. The low-energy spectrum of the spin Hamiltonian
can be interpreted in the language of (quasi)particles: a finite energy
gap corresponds to a massive particle at rest, and the dispersion
relation of the excitation for small momenta can be read as a relativistic
(on shell) energy-momentum relation. The field-theoretic approach
proved to be extremely powerful to explain experimental data. For
instance, recent observations on spin-1/2 compounds with Dzyaloshinskii-Moriya
interaction in external field \cite{asano2000,kenzelmann2004,zyagin2004},
confirm the appearence of effective particles, namely solitons and
breathers, as predicted by the Sine-Gordon Model (SGM) that describes
the low energy spectrum of the chain. 

In this paper we discuss the possibility of observing solitons and
possibly breathers in spin-1 chains with internal anisotropies. In
particular, it is important to understand the role of the interactions
between the particles of the continuum theory to see how their bound
and scattering states manifest in the energy spectrum of the lattice
model. The effective interactions will depend on the values of parameters
of the spin Hamiltonian and it is possible that certain two- (or more)
particle states that are stable in a region of parameter space loose
their stability when the parameters are continuously changed. In other
terms the formerly stable state acquires a rest mass equal or larger
than the sum of the rest masses of its constituents and, if there
are no special selection rules, the particle decays into the continuum
part of the spectrum. 

In this paper we study such a scenario for the following spin-1 anisotropic
model:\begin{equation} 
H=\sum_{j=1}^{L}\left\{ S_{j}^{x}S_{j+1}^{x}+S_{j}^{y}S_{j+1}^{y}+\lambda S_{j}^{z}S_{j+1}^{z}+D\left(S_{j}^{z}\right)^{2}\right\} ,\label{hamilt}\end{equation}
 where $\lambda$ and $D$ parametrize the Ising-like and single-ion
anisotropies respectively. This Hamiltonian is known to be described,
in the continuum low-energy limit, by a Non-Linear Sigma Model (NL$\sigma$M)
in the vicinity of the isotropic antiferromagnetic Heisenberg point
($\lambda=1,\, D=0$) \cite{haldane83,aue}, and the SGM in the neighborhood
of the critical line separating two gapped phases, namely the large-$D$
and the Haldane ones \cite{on_c1_line}. These phases will be defined
in section \ref{sec:QFT}, where we will also recall the mapping onto
the SGM and present a novel extension of the NL$\sigma$M for anisotropic
integer-spin models which encompasses the usual isotropic NL$\sigma$M
that lies at the basis of the Haldane conjecture. In general we find
a triplet of excitations and no other bound states. On the contrary,
from a quantitative analysis of the SGM we expect that in some region
of the Haldane and large-$D$ phases at least an additional bound
state (a breather) should appear. We proceed in section \ref{sec:Num}
to a numerical check of these expectations using the multi-target
Density Matrix Renormalization Group (DMRG) technique that allows
us to handle chains of up to 100 sites and extract various excited
states of the spectrum. The data in the large-$D$ phase confirm the
existence of a breather in the region predicted by the SGM. On the
other hand, we do not find any stable breather in the Haldane phase.
In section \ref{sec:PTLD} we will sketch also a simple perturbative
argument that provides a quantitative interpretation of the data.
Finally, in section \ref{sec:Conc} we will comment on our results
and draw some general conclusions.

\section{Quantum field theories for the $\lambda-D$ model\label{sec:QFT}}

\begin{figure}
\begin{center}\includegraphics[%
  width=8cm,
  keepaspectratio,clip]{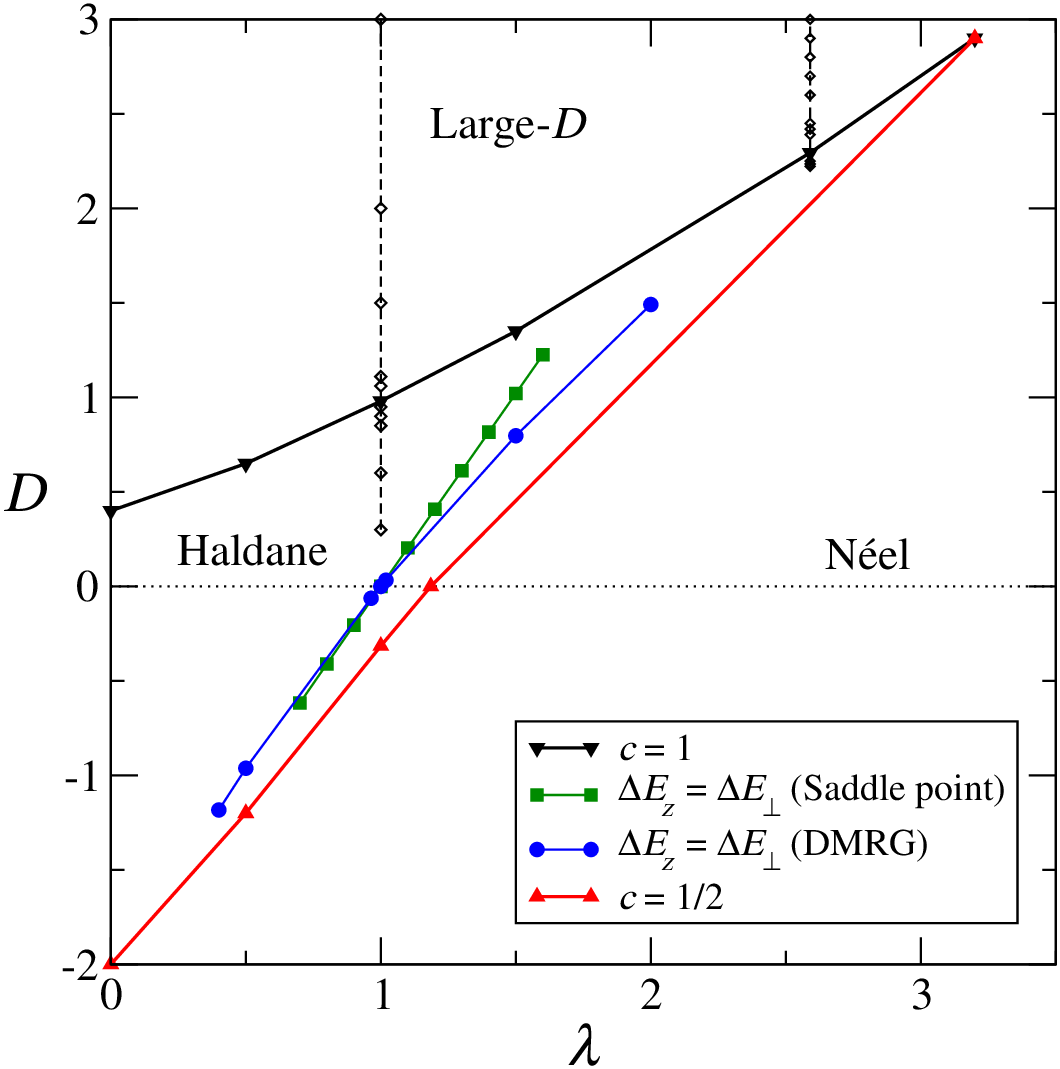}\end{center}

\caption{\label{fig:p-d}Phase diagram of the model (\ref{hamilt}) obtained
numerically via DMRG. The two lines passing through the isotropic
point are the loci of points where the transverse and longitudinal
gaps coincide. Circles are DMRG data whereas squares are the results
of the theoretical approach. The vertical lines across the Haldane-large-$D$
transitions join the points studied numerically (see section \ref{sec:Num}). }
\end{figure}

Among integer spin models, the anisotropic $S=1$ chain with its rich
phase diagram occupies a relevant position. At the isotropic
point, ($\lambda=1$, $D=0$) the ${\rm O}(3)$ Heisenberg model is
recovered. The full phase diagram consists of six different phases
(see \cite{chs} for a recent numerical determination). We will focus
our attention on the $\lambda>0$ half-plane where, apart from the
transition lines, all the phases show a nonzero energy gap above the
ground state (GS). In fig.~\ref{fig:p-d} we report the phase diagram
in this range of parameters. Usually, two different types of order
parameters are used to characterize these phases: the N\'{e}el order
parameters (NOP):

\begin{equation}
{\cal O}_{N}^{\alpha}=\lim_{\vert i-j\vert\rightarrow\infty}(-1)^{i-j}\langle S_{i}^{\alpha}S_{j}^{\alpha}\rangle;\;\alpha=x,y,z\label{nop}\end{equation}
and the string order parameters (SOP):

\begin{equation}
{\cal O}_{S}^{\alpha}=-\lim_{\vert i-j\vert\rightarrow\infty}\langle S_{i}^{\alpha}{\rm e}^{{\rm i}\pi\sum_{k=i+1}^{j-1}S_{k}^{\alpha}}S_{j}^{\alpha}\rangle;\;\alpha=x,y,z\,,\label{sop}\end{equation}
first introduced by den Nijs and Rommelse \cite{nijs_rom}. For $D\gg1$
and $D\gtrsim\lambda$ 
the system is in the so-called \textit{large-$D$ phase}, with a unique
GS that does not break the above symmetry:
$\mathcal{O}_{S}^{\alpha}=\mathcal{O}_{N}^{\alpha}=0$ for all $\alpha$.
Decreasing $D$, for $\lambda\lesssim3$, we have a transition into
the \textit{Haldane phase}. This phase is characterized by the non-vanishing
of all the components of the SOP's $\mathcal{O}_{S}^{\alpha}\neq0$,
meaning that the ${\rm Z}_{2}\times{\rm Z}_{2}$ symmetry is fully
broken, and by $\mathcal{O}_{N}^{\alpha}=0\,\,\forall\alpha$. Inside
the Haldane phase, the gap of the first excited state, though always
different from zero, belongs to two different spin sectors depending
on the parameters $\lambda$ and $D$. It belongs to the $S_{{\rm tot}}^{z}=\pm1$
sector for values of $\lambda$ and $D$ above the so-called degeneracy
line (see fig.~\ref{fig:p-d}). This line, which passes through the
${\rm O}(3)$ isotropic point, divides the Haldane phase into two
sub-phases \cite{botet}. Below this line the first excited state
has $S_{{\rm tot}}^{z}=0$. The GS is always unique and belongs to
the $S_{{\rm tot}}^{z}=0$ sector. On the right of the Haldane phase
we find the twofold degenerate \textit{N\'{e}el phase}. Here the
${\rm Z_{2}}\times{\rm Z}_{2}$ symmetry is only partly broken as
as proved by the fact that $\mathcal{O}_{N}^{\alpha}=\mathcal{O}_{S}^{\alpha}=0$
for $\alpha=x,y$ but $\mathcal{O}_{N}^{z},\,\mathcal{O}_{S}^{z}\neq0$.
The Haldane-large-$D$ (H-D) and Haldane-\textit{\emph{N\'{e}el (H-N)}}
transition lines mark second order transitions; the former is described
by a $c=1$ Conformal Field Theory (CFT) while the latter by a $c=1/2$
CFT. They merge at the tricritical point ($\lambda\cong2.90$, $D\cong3.20$)
\cite{on_c1_line} where the Haldane phase disappears and the large-$D$-N\'{e}el
transition becomes first order.

Near the isotropic point the Hamiltonian (\ref{hamilt}) can be mapped
onto an anisotropic version of the NL$\sigma$M using Haldane's ansatz
\cite{aue} to decompose the spin coherent states field $\vec{\Omega}_{j}(\tau)=\vec{n}_{j}(\tau)(-1)^{j}\sqrt{1-\frac{\ell_{j}^{2}(\tau)}{S^{2}}}+\frac{\vec{\ell_{j}}(\tau)}{S}$
into a uniform, $\vec{\ell_{j}}(\tau)$, and a staggered part $\vec{n}_{j}(\tau)=[\mathbf{n}_{j\perp}(\tau),n_{jz}(\tau)]$,
satisfying the constraint $|\mathbf{n}_{\perp}|^{2}+n_{z}^{2}=1$. After
integration of the uniform field, one obtains the following 
effective Lagrangian density:\begin{eqnarray}
\mathcal{L} & = & \frac{v_{\perp}}{2g_{\perp}}\left[\vert\partial_{x}{\bf \mathbf{n}}_{\perp}\vert^{2}+\frac{1}{v_{\perp}^{2}}\vert\partial_{\tau}\mathbf{n}_{\perp}\vert^{2}\right]+\nonumber \\
 &  & \frac{v_{z}}{2g_{z}}\left[\left(\partial_{x}n_{z}\right)^{2}+\frac{1}{v_{z}^{2}}\left(\partial_{\tau}n_{z}\right)^{2}+\mu_{z}n_{z}^{2}\right]+\mathcal{L}_{\textrm{int}},\label{eq:effective_L}\end{eqnarray}
where the interaction term has a simple expression only in the  
regions $|n_z| \ll  1$ and $| \mathbf{n}_\perp| \ll 1$. In the first region
the interaction term has the form\[
\mathcal{L}_{\textrm{int}}=n_{z}^{2}\left[d_{\perp}\left|\partial_{\tau}\mathbf{n}_{\perp}\right|^{2}+d_{z}(\partial_{\tau}n_{z})^{2}\right],\]
 whereas, in the region $\left|{\bf \mathbf{n}}_{\perp}\right|\ll1$,
the interaction term has the form \[
\mathcal{L}_{\textrm{int}}=\left|\mathbf{n}_{\perp}\right|^{2}\left[d_{\perp}\left|\partial_{\tau}\mathbf{n}_{\perp}\right|^{2}+d_{z}(\partial_{\tau}n_{z})^{2}\right].\]
 Here the constants $g_{\perp\left(z\right)},\, v_{\perp\left(z\right)},\, d_{\perp\left(z\right)},\,\mu_{z}$
are different functions of the microscopic parameters $\lambda$ and
$D$ whose explicit expressions depend on the region one considers.

As long as one deals with integer spins the topological term arising
from the combination of the Berry phases on each site reduces to a
multiple of $2\pi$ and therefore can be safely omitted. More details
on the mapping into this anisotropic variant of the NL$\sigma$M can
be found in ref.~\cite{pas}. 

The constraint of unit norm makes the theory hard to treat. In both
regions, $\left|n_{z}\right|\ll1$ and $\left|{\bf \mathbf{n}}_{\perp}\right|\ll1$,
we limit ourselves to a mean-field solution that can be worked out
using the saddle-point approximation on eq.~(\ref{eq:effective_L}).
The constraint is taken into account by introducing a uniform Lagrange
multiplier $\eta$. We refer the reader to ref.~\cite{pas} for details
and only mention that one arrives at a couple of self-consistent equations
for the longitudinal and transverse gaps $\Delta E_{z}$ and $\Delta E_{{\rm \perp}}$,
that play the role of the masses of the particles of an {}``anisotropic
Haldane triplet''. Of course $\Delta E_{z}=\Delta E_{\perp}$ at
the isotropic Heisenberg point. However, when $\lambda$ and $D$
are varied the transverse and longitudinal channels split. Nonetheless,
guided by the numerical work of \cite{botet}, we can search for a
specific line in parameter space where the two gaps remain degenerate
(in the thermodynamic limit) even if the lattice theory has no ${\rm O}(3)$
symmetry. Interestingly, the line found through the self-consistent
equations is in good agreement with the one found with the DMRG as
depicted in fig.~\ref{fig:p-d}. 

Let us now see a possible connection with the SGM. Neglecting the
longitudinal field (i.e.~setting $n_{z}=0$) and writing $\mathbf{n}_{\perp}=\mathrm{e}^{i\theta}$,
then an ${\rm O}(2)$ NL$\sigma$M is recovered:

\begin{equation}
\mathcal{L}=\frac{1}{2}v\left[\left(\partial_{x}\Theta\right)^{2}+\frac{1}{v^{2}}\left(\partial_{\tau}\Theta\right)^{2}\right]\label{O(2)_NLSM}\end{equation}
with $\Theta=\theta/g$, $g=\sqrt{(2+2\lambda+D)}$ and $v=g$ (for
$S=1$) %
\footnote{In ref. \cite{on_c1_line} in all the expressions after eq. (6) $D$
has to be replaced with $D/2$. This replacement affects the theoretical
values of $K$ and of the scaling dimensions, but not their numerical
estimates. %
}. This is exactly a free Gaussian model with a bosonic field compactified
along a circle of radius $1/\sqrt{g}$. We know that this model describes
a CFT with central charge $c=1$ and with primary fields having scaling
dimensions: 

\begin{equation}
d_{mn}=\left(\frac{m^{2}}{4K}+n^{2}K\right),\label{scalDimGauss}\end{equation}
with $K=\pi/g$ and $m,n\in\mathbb{Z}$.

In a previous work \cite{on_c1_line} it was argued that near the
Gaussian $c=1$ line the perturbed ${\rm O}(2)$ model can written
in terms of a sine-Gordon model:

\begin{equation}
{\cal L}=\frac{1}{2}\left[v\left(\partial_{x}\Phi\right)^{2}+\frac{1}{v}\left(\partial_{\tau}\Phi\right)^{2}\right]+\frac{v\bar{\mu}}{a^{2}}\cos(\sqrt{4\pi K}\Phi)\label{SG}\end{equation}
where $\Phi$ is a compactified field dual to $\Theta$, $a$ is the
lattice spacing and $\bar{\mu}$ is a coupling constant that should
vanish exactly along the H-D transition line. $K$ is the basic parameter
that allows to compute all the scaling dimensions, and consequently
all the critical exponents, that along the $c=1$ line acquire nonuniversal
values depending on the actual values of $\lambda$ and $D$: it ranges
from the value of $K=2$ at the Berezinskii-Kosterlitz-Thouless (BKT)
point ($\lambda=0$, $D\simeq0.4$) to the value $K=1/2$ at the tricritical
point, becoming $K=1$ at the so-called free Dirac point, $\lambda\simeq2$.
The SGM has a spectrum consisting of a soliton, an anti-soliton and
their bound states, breathers, the number of which is different from
zero only for $K\le1$ \cite{lz}. This means that, besides the excitations
of the $c=1/2$ theory (belonging to $S_{{\rm tot}}^{z}=0$), in the
region of the phase diagram that remains on the left side of the free
Dirac point (i.e.~approximately for $0<\lambda<2$, for which $1<K<2$)
the particle content of the theory should consist only of a soliton
and an anti-soliton. When $K\leq1$ (i.e.~$2<\lambda<2.90$), also
a breather should appear (and possibly a second breather for $K\geq2/3$).
If this is proved to be correct, then it would mean that the Haldane
and the large-$D$ phases could be eventually classified into subphases
according to the number of stable particles. In order to check this
hypothesis, we have numerically analyzed the spectrum of the model
(\ref{hamilt}) on the $\lambda=1$ and $\lambda=2.59$ lines for
values of $D$ crossing the Haldane-large-$D$ transition curve.

\section{Numerical investigation\label{sec:Num}}

\subsection{$K>1$: No breathers}

For $\lambda=1$ the critical point on the $c=1$ line has been previously
located at $D_{{\rm c}}=0.99$ with a parameter $K=1.328\pm0.004$
\cite{on_c1_line}; then we are in the sector of the Haldane phase
that should be characterized only by a soliton and an anti-soliton.
From CFT we know that a free Gaussian theory for a field $\Theta$
compactified along a circle with radius $\sqrt{K/\pi}$ has primary
fields of scaling dimensions given by equation (\ref{scalDimGauss})
where $m$ plays the role of $S_{{\rm tot}}^{z}$. The energies of
the excited states are related to the scaling dimensions by the formula:

\begin{equation}
\Delta E_{mn}=E_{mn}-E_{{\rm GS}}=\frac{2\pi v}{L}\left(d_{mn}+r+\bar{r}\right),\; r,\bar{r}\in\mathbb{N}.\label{CFT_excited}\end{equation}
The spectrum of the scaling dimension is reported in table 2 of ref.
\cite{on_c1_line}. In the language of the SGM, labelling the states
as $\left(m,n,r,\bar{r}\right)$, the first excited states are given
by $(\pm1,0,0,0)$ and belong to the $S_{{\rm tot}}^{z}=\pm1$ spin
sectors. They correspond to the soliton and the anti-soliton, while
the first breather comes from the doublet $(0,\pm1,0,0)$ that splits
into two different states as soon as one moves away from criticality.
Operatively, a breather is a singlet in the $S_{{\rm tot}}^{z}=0$
spin sector. To be a stable state, the ratio between its rest energy
and that of the soliton (measured respect to the ground state) $R=\Delta E_{b}/\Delta E_{s}$
must be smaller than two.

The numerical analysis has been performed with a DMRG algorithm that
exploits the so-called thick-restart Lanczos method (see \cite{sch}
for a recent review and \cite{debo} for our implementation). We adopt
periodic boundary conditions (PBC) to avoid complicancies arising
from edge effects (midgap states and surface contributions see
e.g.~\cite{wang2000}), and perform 
up to five finite-system iterations in order to reduce the uncertainty
on the energies to the order of magnitude of the truncation error
\cite{lf}. Using $M=400$ DMRG states the latter is $O(10^{-4})$
or better, the worst cases being the ones with 10 target states. 

\begin{figure}
\begin{center}\includegraphics[%
  width=8cm,
  keepaspectratio,
clip]{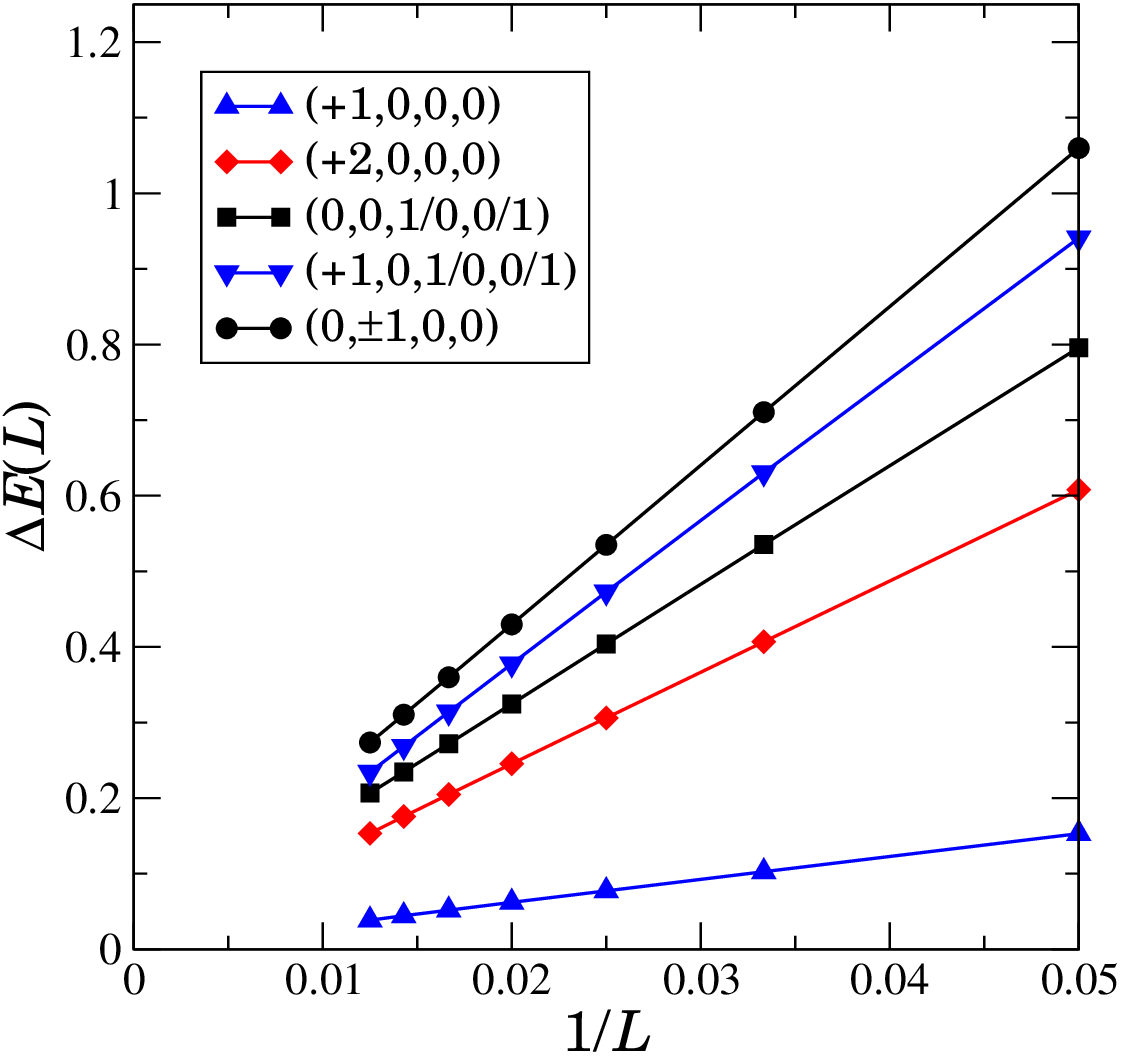}\end{center}

\caption{Plots of the low-lying energy gaps versus $1/L$ for $D=0.90$ $\lambda=1$.
The states are labeled with the quantum numbers of the CFT (see text).
\label{fig:quasi-critical} }
\end{figure}

On the $\lambda=1$ line we have studied points close the Haldane-large-$D$
transition as well as points inside the gapped Haldane and large-$D$
phases. For $D=0.85$, $0.90$, $0.95$, $1.06$, $1.11$, we are still in
a quasi-critical region for which the correlation length is still
larger than the size of the system. This is confirmed by the fact
that the gaps scale substantially linearly in $1/L$ as predicted
by eq.~(\ref{CFT_excited}), and we can label the states with the
quantum numbers of the CFT at the critical point. A typical situation
is shown in fig.~\ref{fig:quasi-critical} and one sees that the
states corresponding to $\left(0,\pm1,0,0\right)$ are still degenerate.
Following the evolution of this doublet for values of $D$ farther
from the critical line, we can identify the first breather as the
lowest of the two splitted states. 

\begin{figure}
\begin{center}\includegraphics[%
  width=8cm,
  keepaspectratio,
clip]{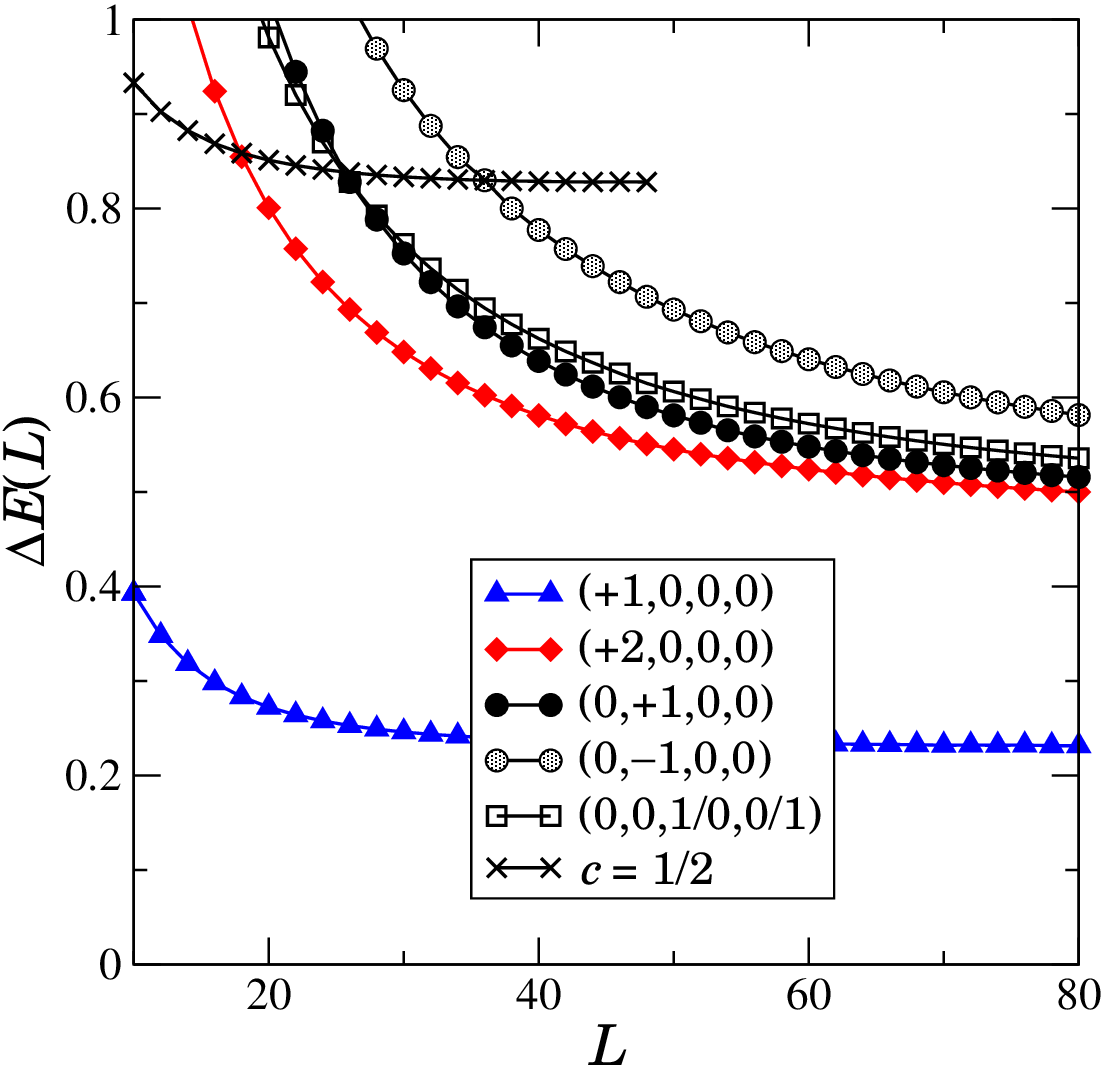}\end{center}

\caption{Plots of the low-lying energy gaps versus $L$ for $D=0.3$ at $\lambda=1$.
States are still labeled as in fig. \ref{fig:quasi-critical}, except
that here the states $(0,\pm1,0,0)$ have split and grey circles represent
the highest of the two. It is already possible to see the first excitation
of the $c=1/2$ theory (crosses); this state appears incomplete because
for $L\ge50$ it lies outside the set of the DMRG-targeted states.\label{off-critical-D0.3}}
\end{figure}

Moving downward with $D$ well into the Haldane phase up to $D=0.3,\,0.6$,
in the $S_{\textrm{tot}}^{z}=0$ sector we observe besides a doublet
state that we identify with the secondary states $\left(0,0,1/0,0/1\right)$,
two clearly separated states that are the evolution of the $(0,\pm1,0,0)$
doublet. This feature is evident in fig.~\ref{off-critical-D0.3}
reporting the case $D=0.3$. At these values of $D$ the correlation
length is sufficiently small, and we extrapolated the asymptotic values
of the energy gap for both the soliton and the breather using the
semi-phenomenological formula \cite{japar}\begin{equation}
\Delta E(L)=\Delta E(\infty)+\frac{A}{L}\exp\left(-\frac{L}{\xi'}\right),\label{eq:fitting_f}\end{equation}
where the fitting parameter $\xi'$ is expected to be proportional
to the actual correlation length. We verified that eq.~(\ref{eq:fitting_f}),
fits better the data than other functions with different exponents
of $1/L$. The extrapolated values for the breather gaps lay slightly
above the continuum threshold, specifically we obtain $R=2.048$ at
$D=0.6$ and $R=2.043$ at $D=0.3$. In addition we verified that
also the lowest state in the $S_{\textrm{tot}}^{z}=2$ sector remains
in the continuum in the thermodynamic limit.

We pushed the analysis a little further for $D\ge1.5$. As the correlation
length is now sufficiently small we trust the data at $L=100$ without
the help of any extrapolations. The result is that also in this regime
$R>2$ so that there are no stable breathers. In agreement with what
observed in \cite{feverati1998a,feverati1998b}, where the finite-size
spectrum of the SGM is studied, the first excitation in the sector
$S_{{\rm tot}}^{z}=0$ lies always above the soliton.

In summary, as far as the $K>1$ region is concerned, the numerical
investigation allows to conclude that there are no soliton bound states
both in the Haldane and in the large-$D$ phases.

\subsection{$K<1$: Emergence of the breather}

Let's analyze now what happens on the $\lambda=2.59$ line, for which
we have previously checked that $D_{{\rm c}}=2.30$ and $K=0.85$
\cite{on_c1_line}. If we believe in the SGM as a faithful continuum
theory of the low energy part of the model (\ref{hamilt}) we expect
the spectrum to present a stable breather both in the Haldane and
in the large-$D$ phases. According to the estimated $K$ the order
of the CFT energy levels is the one reported in table \ref{tab_Scaldim2}
where we have shown the scaling dimensions $d^{\textrm{CFT}}=d_{m n}+r+\bar{r}$.

\begin{table}
\begin{center}\begin{tabular}{|l|c|}
\hline 
$d^{{\rm CFT}}${[}$\times$ degeneracy{]}&
 $(m,n,r,\bar{r})$\tabularnewline
\hline
\hline
0 {[}$\times$1{]}&
 (0,0,0,0)\tabularnewline
\hline
$0.294\pm0.003$ {[}$\times$2{]}&
($\pm1$,0,0,0)\tabularnewline
\hline
$0.85\pm0.01$ {[}$\times$2{]}&
(0,$\pm1$,0,0)\tabularnewline
\hline
1 {[}$\times$2{]}&
(0,0,1/0,0/1)\tabularnewline
\hline 
$1.18\pm0.01$ {[}$\times$2{]}&
($\pm2$,0,0,0)\tabularnewline
\hline
$1.29\pm0.02$ {[}$\times$4{]}&
($\pm1$,0,1/0,0/1)\tabularnewline
\hline
\end{tabular}\end{center}

\caption{Spectrum of scaling dimensions at the point ($\lambda=2.59$,$D_{{\rm c}}=2.30$),
obtained from formula (\ref{scalDimGauss}) where $K=0.85\pm0.01$
\cite{on_c1_line}.\label{tab_Scaldim2}}
\end{table}

As in the case $K>1$, we are interested in the soliton gap originating
from the $\left(\pm1,0,0,0\right)$ states and in the breather coming
from the $\left(0,\pm1,0,0\right)$ states.

The correlation length now decreases rather rapidly with $\left|D-D_{c}\right|$
so that we see the off-critical regime for $D$ quite close to the
critical point for the system sizes at our disposal. 

\begin{figure}
\begin{center}\includegraphics[%
  width=8cm,
  keepaspectratio,clip]{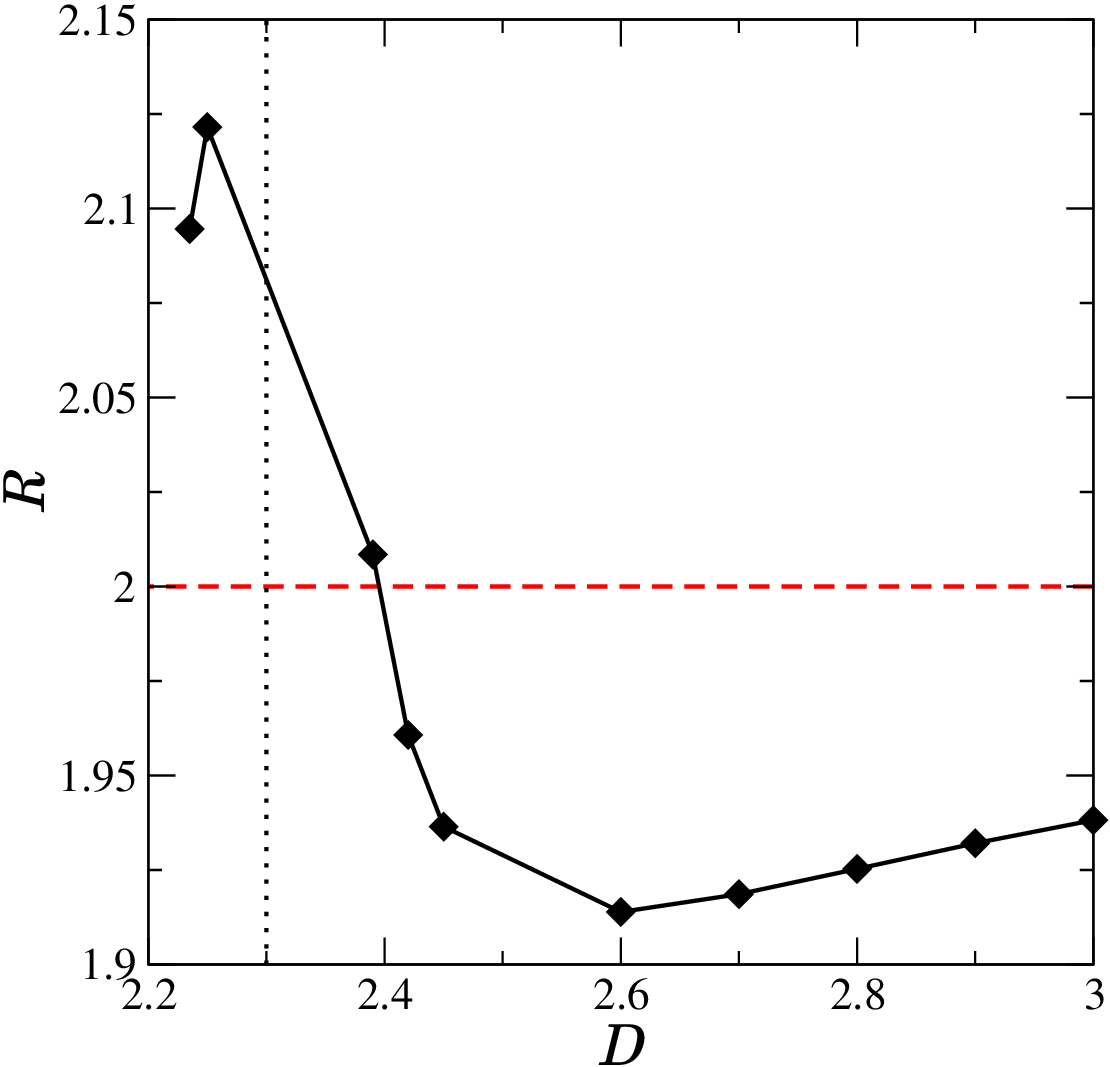}\end{center}

\caption{Plot of the ratio $R$ between the breather and the soliton energy
gaps at $L=100$ as a function of $D$ for $\lambda=2.59$. The dashed
horizontal line represents the continuum threshold while the dotted
vertical line marks the critical point. \label{fig:ratio}}
\end{figure}

Inside the large-$D$ phase we selected eight points with $D$ ranging
from 2.39 to 3. We have checked that the correlation length is indeed
so small that we can use the data at $L=100$ without any finite-size
scaling. The identification of the soliton and the breather states
out of the DMRG spectrum is rather direct. Apart from a single point
quite close to the critical line, the gap ratio $R$ is always less
than two, as shown in fig.~\ref{fig:ratio}. This confirms the existence
of an additional stable particle that we identify as the breather.
Considering the non-monotonic behavior of the function $R\left(D\right)$,
we can speculate whether the breather looses its stability for very
large values of $D$. In the next section we will study this point
by means of an analytical approach. 

\begin{figure}
\begin{center}\includegraphics[%
  width=8cm,
  keepaspectratio,clip]{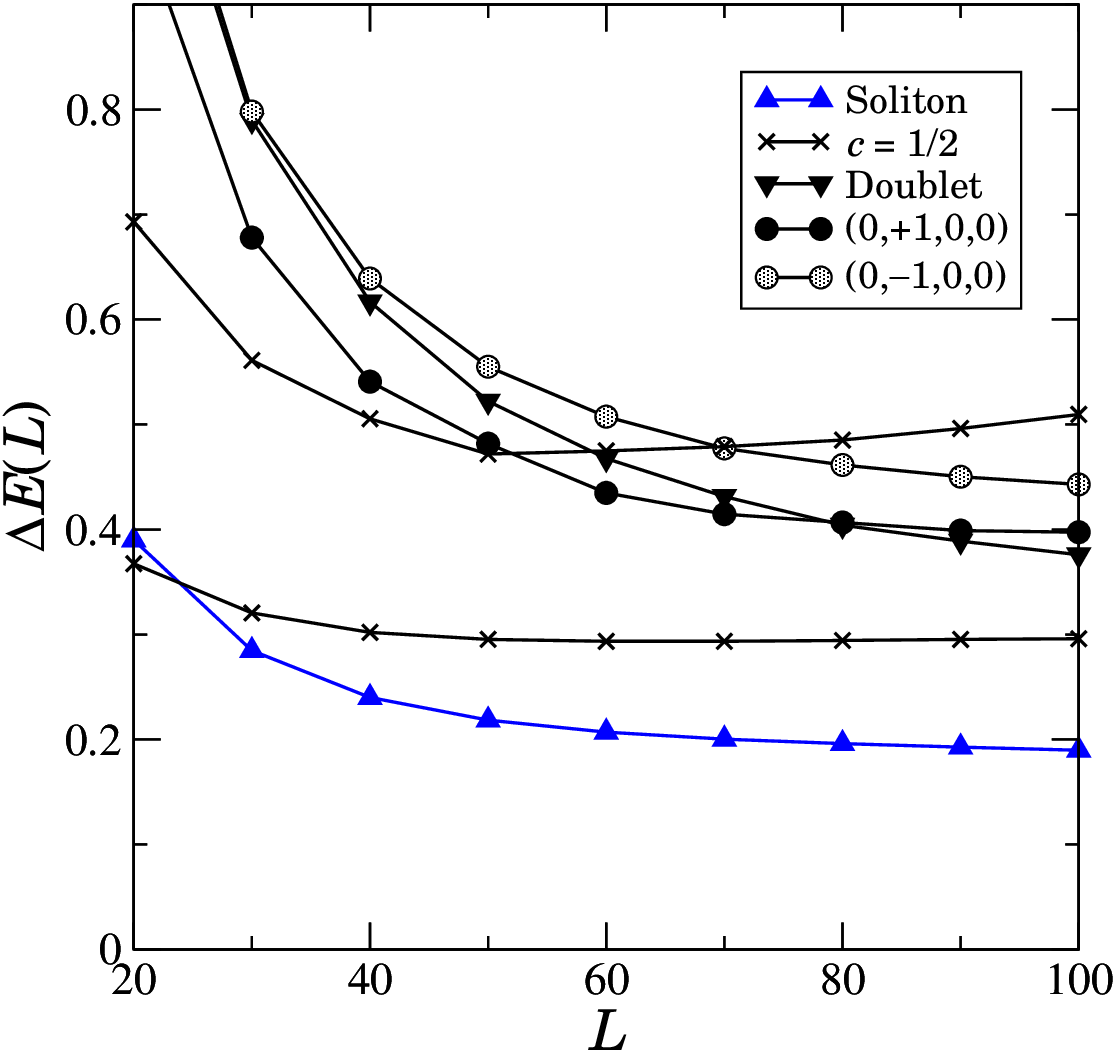}\end{center}

\caption{Typical low-lying spectrum in the Haldane phase for $K<1$ ($\lambda=2.59,\, D=2.235$).
\label{fig:lambda2.59_typical}}
\end{figure}

In the Haldane phase, which is very narrow here, we studied the points
$D=2.2222,\,2.235,\,2.25$. The identification of the soliton and
the breather is now complicated by the appearance of states which
do not originate from the $c=1$ CFT. A typical numerical spectrum
up to $L=100$ is shown in fig.~\ref{fig:lambda2.59_typical}. Following
the states in the $S_{\textrm{tot}}^{z}=0$ sector, labelled by crosses,
for smaller values of $D$ we observe that their gaps decrease and
eventually vanish at the $c=1/2$ transition line whereas all the
other states remain massive. For this reason we identify these states
as excitations coming from the $c=1/2$ CFT. The doublet (triangles
down) cannot be identified with the breather being twofold degenerate;
it is probably a descendent ($r=0/1,\,\bar{r}=1/0$) of the first
$c=1/2$ state. What we find is that, both at $D=2.25$ and at $D=2.235$,
the breather lies above the two-soliton gap continuum, i.e.~$R>2$.
The point $D=2.2222$ lies on the {}``degeneracy-line'' where the
soliton gap is degenerate with the level originating from the first
excitation of the $c=1/2$ theory. Now many states fall below the
breather so that the latter can no longer be seen within the numerically
available levels. 

\section{Perturbation theory at large-$D$\label{sec:PTLD}}

From the data reported in fig. \ref{fig:ratio} one is led to separate
the large-$D$ phase into two regions depending on the stability of
the breather. However, the location of the curve $\lambda_{{\rm stab}}(D)$
separating the two regions is still vague: it starts from the $c=1$
H-D line at the point where $K=1$, close to $\lambda\simeq2$, but
we have no precise indications for larger values of $D$. In this
region a perturbative analysis can be carried on. From the results
quoted in \cite{orendac1999} it is known that bound states (at zero
total momentum) do not exist for $\lambda=1$. Here we show that the
presence of an Ising-like anisotropy $\lambda$ allows for a stabilization
of breather states. 

We start by treating the $D$-term as unperturbed Hamiltonian 

\begin{equation}
H_{0}\equiv D\sum_{j=1}^{L}\left(S_{j}^{z}\right)^{2},\label{eq:H0}\end{equation}
and the rest as small perturbation\textcolor{black}{\[
V\equiv\sum_{j=1}^{L}\left\{ \frac{1}{2}\left[S_{j}^{+}S_{j+1}^{-}+S_{j}^{-}S_{j+1}^{+}\right]+\lambda S_{j}^{z}S_{j+1}^{z}\right\} .\]
 }

Both $H_{0}$ and $V$ commute separately with the total $z$-component
of the spin, $S_{{\rm tot}}^{z}$, and with the total {}``spin-flip''
(or {}``time-inversion'') operator $T=\exp({\rm i\pi}S_{{\rm tot}}^{y})$.
However, $[S_{{\rm tot}}^{z},T]\neq0$ and one can specify both quantum
numbers for the energy eigenstates only in the sector $S_{{\rm tot}}^{z}=0$.
The GS falls precisely in this sector and here we expect to find the
breather. The soliton and the anti-soliton have $S_{{\rm tot}}^{z}=\pm1$,
but we can restrict to the soliton case $S_{{\rm tot}}^{z}=1$ since
the anti-soliton state is simply obtained by applying $T$. 

\begin{table}
\begin{center}\begin{tabular}{|l|c|c|c|}
\hline 
State&
Energy&
Degeneracy&
Name\tabularnewline
\hline
\hline 
$\vert0_{1}\cdots0_{L}\rangle$&
$0$&
1&
GS\tabularnewline
\hline 
$\vert0_{1}\cdots+_{j}\cdots0_{L}\rangle$&
$D$&
$L$&
Soliton\tabularnewline
$1\le j\le L$&
&
&
\tabularnewline
\hline 
$\vert0_{1}\cdots+_{j}\cdots-_{k}\cdots0_{L}\rangle$&
$2D$&
$L\left(L-1\right)$&
Breather\tabularnewline
$1\le j\neq k\le L$&
&
&
\tabularnewline
\hline
\end{tabular}\end{center}

\caption{Low energy states of the unperturbed Hamiltonian (\ref{eq:H0}).\label{tab:H0-Low-energy-states}}
\end{table}

\begin{figure}
\begin{center}\includegraphics[%
  width=8cm,
  keepaspectratio,clip]{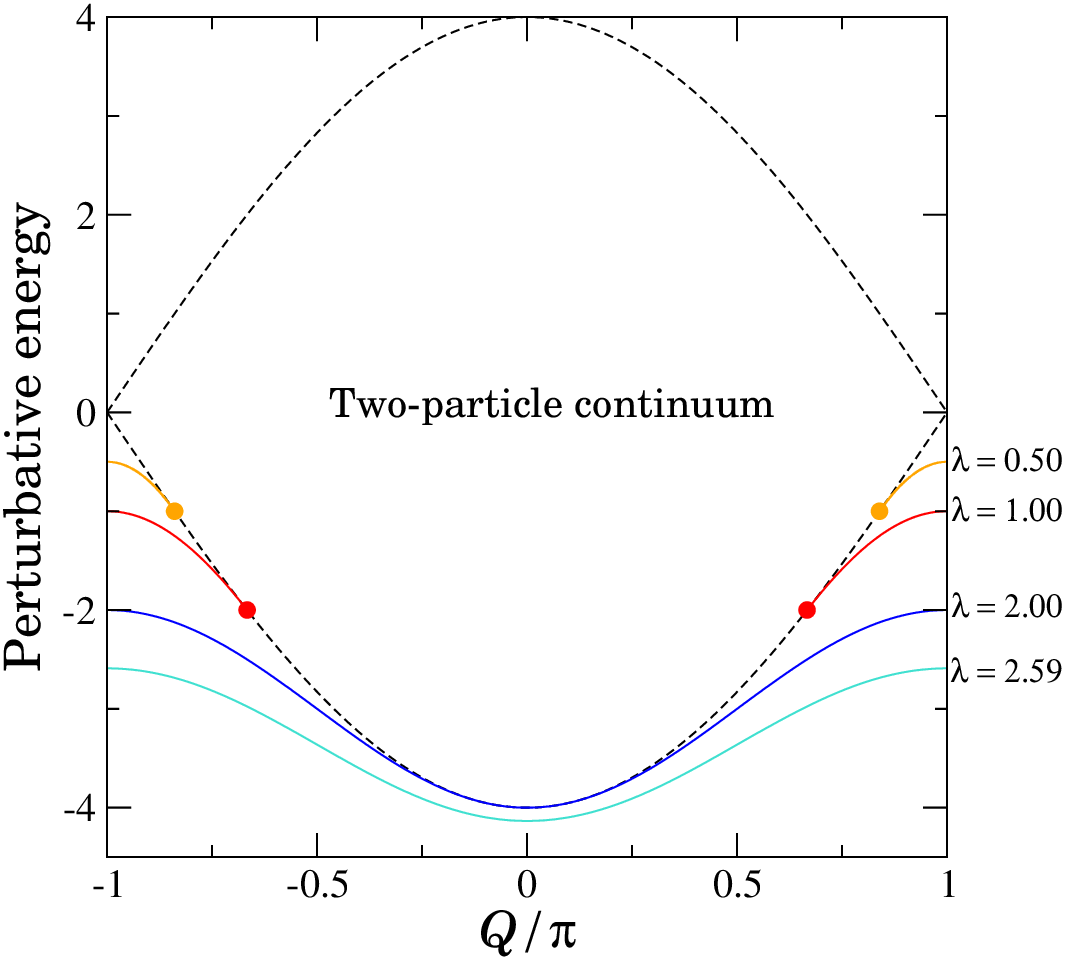}\end{center}

\caption{Bound states (full lines) and continuum (region enclosed by dashed
lines) for the values of $\lambda$ indicated on the right. Full dots
mark the wavenumbers $Q^{*}$ where the bound state starts to detach
from the continuum when $\lambda<2$.\label{fig:two-particles-states} }
\end{figure}

\begin{figure}
\begin{center}\includegraphics[%
  width=8cm,
  keepaspectratio,clip]{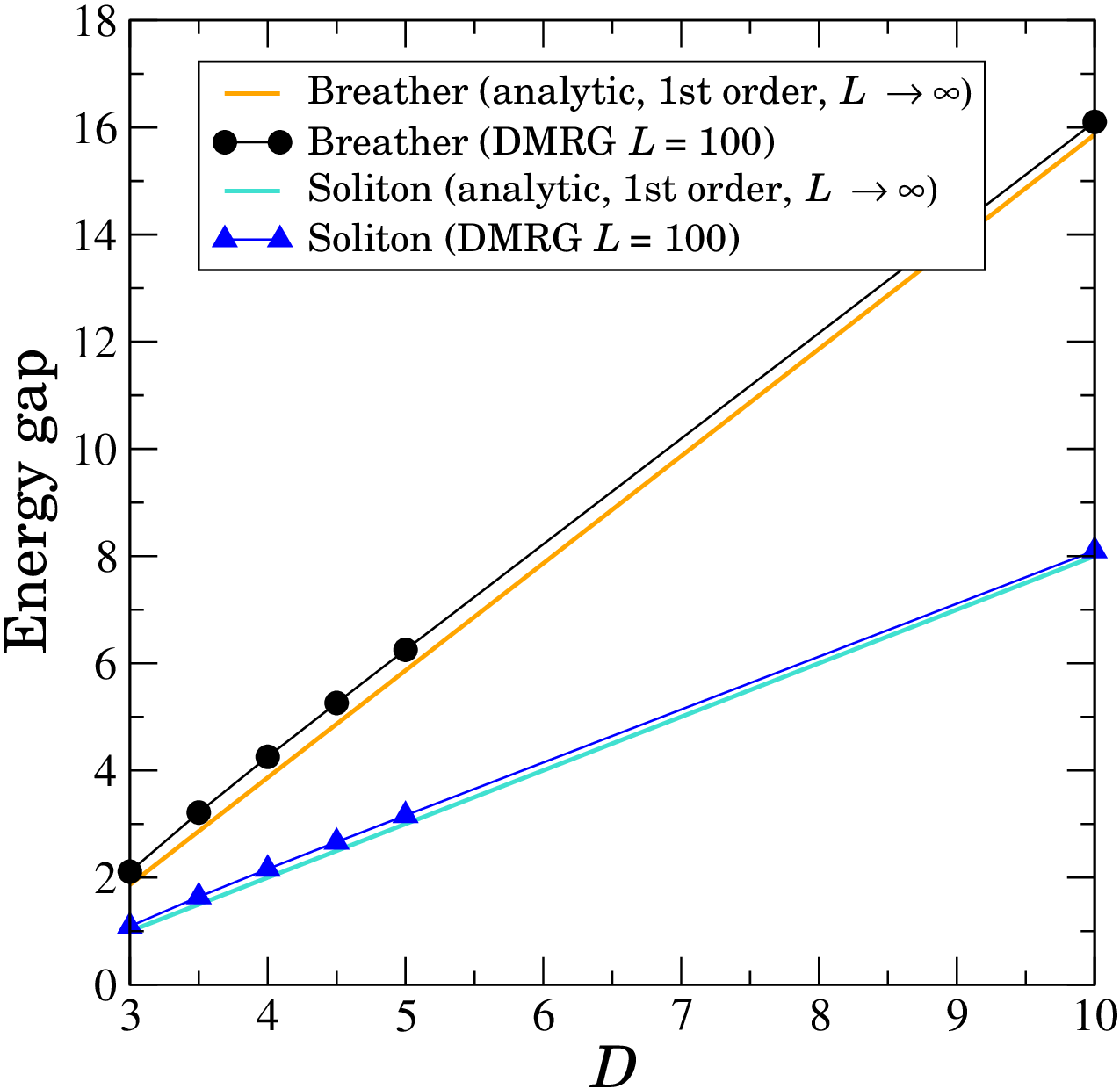}\end{center}

\caption{Results of first-order perturbation theory vs DMRG data in the large-$D$
phase at $\lambda=2.59$. \label{fig:perturbative-gaps} }
\end{figure}

In table \ref{tab:H0-Low-energy-states} we show the GS together with
the first excited states for the unperturbed Hamiltonian (\ref{eq:H0}).
In this picture the ``solitons'' are completely degenerate and
non-interacting so that the (soliton-antisoliton) ``breather's''
energy is exactly twice that of the 1-particle states. As we switch
on the interaction $V$, the GS energy becomes $E^{\left(0\right)}=-L/D+O\left(D^{-2}\right)$
up to second order in perturbation theory. The first excitations are
no longer degenerate; they rather form a band, labelled by the lattice
momentum $q$, $E_{q}^{\left(1\right)}=D+2\cos\left(q\right)+O\left(D^{-1}\right)$.
Up to first order perturbation theory these 1-particle states form
a continuum of two-particle scattering states whose upper and lower
edges are given by the dashed lines in fig.~\ref{fig:two-particles-states}.
As a result of the interaction the soliton states may form bound state
if their energy lies below the two-particle continuum. At first perturbative
order the breather energy can be calculated using the Bethe-Ansatz
approach; details can be found in Appendix A. The resulting breather
states are indexed by the center-of-mass momentum $Q$. Acceptable
states are those for which $\cos\left(Q/2\right)<\lambda/2$. This
condition defines a characteristic wavenumber $Q^{\ast}$ above which
the breather form a bound state with energy \begin{equation}
E_{Q}^{\left(2\right)}=2D-\lambda-\frac{4}{\lambda}\cos\left(Q/2\right)+O\left(D^{-1}\right),\,\textrm{for}\,\,\left|Q\right|>Q^{\ast}.\label{eq:E2-perturbative}\end{equation}
Note that for $Q=\pi$ the bound state is always below the continuum
with an energy shift $\delta E_{\pi}^{\left(2\right)}=-\lambda$. However
with the DMRG we inspect the low-lying part of the spectrum, and in
order to find a stable bound state, that will be identified with the
breather, we must examine the minimum of the continuum at $Q=0$.
In this case the bound states emerges from the continuum only for
$\lambda>2$ with an energy shift $\delta E_{0}^{\left(2\right)}=-\lambda-4/\lambda$.
The mechanism is illustrated in fig. \ref{fig:two-particles-states}
for some values of $\lambda$. Interestingly enough, the mass ratio
$R=\Delta E_{{\rm b}}(\infty)/\Delta E_{{\rm s}}(\infty)$ now reads:\begin{equation}
R=\frac{E_{0}^{\left(2\right)}}{E_{\pi}^{\left(1\right)}}=2-\frac{1}{\lambda}\frac{\left(\lambda-2\right)^{2}}{\left(D-2\right)}.\label{RPT}\end{equation}
This equation is valid for $\lambda>2$, meaning that the stability
boundary is $\lambda_{{\rm stab}}(D)=2$ independent of $D$. This
result is consistent with the stability criterion provided by the
SGM: $K=1$ just for $\lambda\sim2$ \cite{on_c1_line}. In addition
the ratio converges to $R=2$ from \textit{below} in agreement with
fig. \ref{fig:ratio}. The first-order approximation is also quantitatively
good in comparison to the DMRG values as shown by the example in fig.
\ref{fig:perturbative-gaps}.


\section{Discussion of the results and conclusions\label{sec:Conc}}

Motivated by field-theoretic predictions, in this paper we have investigated
the stability of massive excitations (soliton, ant-isoliton and breathers)
that are present in the large-$D$ and Haldane phases of the anisotropic
spin-1 Heisenberg chain with both Ising-like and single-ion anisotropy.

In excellent agreement with the SGM that is supposed to describe the
continum limit of the lattice model in the vicinity of the H-D critical
line, we have provided both numerical and analytical results supporting
the existence of a stable breather in the large-$D$ phase of the
model (\ref{hamilt}) in the region with $\lambda>2$.

On the contrary, as far as the Haldane phase is concerned, our best
estimates do not show other stable states in addition to the three
particles that form the Haldane triplet at the isotropic Heisenberg
point. This is at variance with what we would expect from the SGM,
indicating that the latter is probably not sufficient to describe
the effective interaction when there is an underlying topological
order as the one found in the Haldane phase \cite{kennedy92}. Indeed
in this phase we may expect that we cannot neglect the effects of
an interaction between the states coming from the $c=1$ CFT describing
the H-D critical line and those pertaining to the $c=1/2$ CFT describing
the H-I transition.

It is interesting to speculate whether this excitation can be observed
in real materials. The so called NENC has the advantage of having
a large single-ion anisotropy $D\simeq7.5$ but $\lambda=1$ in this
compound. On the other hand large Ising anisotropies are present in
some spin-1/2 materials (e.g.~$\textrm{CsCoCl}_{3}$ and $\textrm{CsCoBr}_{3}$;
see \cite{asano2002} and references therein). Interestingly enough,
differently from existing experiments the mechanism proposed here
to observe breather states does not require the presence of external
magnetic fields.

\begin{acknowledgement}
This work was supported by the TMR network EUCLID Contract No. HPRN-CT-2002-00325,
and the COFIN projects, Contracts Nos. 2002024522001 and 2003029498013.
\end{acknowledgement}
\appendix

\section*{Appendix A}

In this appendix we show how to evaluate the first order correction
of the second energy level in $S_{\textrm{tot}}^{z}=0$. The unperturbed
states are\[
\vert u_{j,k}^{\left(2\right)}\rangle=\vert0_{1}\cdots+_{j}\cdots-_{k}\cdots0_{L}\rangle,\,\, j\neq k\]
 and we need the eigenvalues of the matrix $\langle u_{l,m}^{\left(2\right)}|V\vert u_{j,k}^{\left(2\right)}\rangle$
with $j\neq k$ and $l\neq m$. We write the --unnormalized-- eigenfunction
as\[
|\psi^{\left(2\right)}\rangle=\sum_{j,k}f_{j,k}\vert u_{j,k}^{\left(2\right)}\rangle.\]
The states can be classified according to the eigenvalue of $T$:\[
T\vert\psi^{\left(2\right)}\rangle=\tau\vert\psi^{\left(2\right)}\rangle,\;\tau=\pm1\;\Rightarrow f_{kj}=\tau f_{jk}.\]
In terms of the coefficients $f_{jk}$ the eigenvalue equation reads:\begin{multline}
(1-\delta_{jk})\left[f_{j+1,k}+f_{j,k-1}+f_{j-1,k}+f_{j,k+1}\right]\\
-\lambda f_{jk}\left(\delta_{j,k+1}+\delta_{j,k-1}\right)=\delta E^{\left(2\right)}f_{jk}.\label{evef}\end{multline}
Note that we have put the term $(1-\delta_{jk})$ in the left-hand
side for compatibility with the requirement $f_{jj}=0$. Now, eq.
(\ref{evef}) is an eigenvalue equation in a $L(L-1)$-dimensional
space. It is actually much more simple to express the amplitudes using
relative and center-of-mass coordinates: $r=j-k$ and $\rho=(j+k)/2$.
From now on we will follow closely the approach used in Mattis' book
\cite{mat} (sec. 5.3) for the bound states of magnons in Heisenberg
ferromagnets. Consider the wave function:\begin{equation}
f_{jk}=\exp\left({\rm iQ\rho}\right)F(r),\; F(r)=\sum_{p}f(p)\exp\left({\rm i}pr\right),\label{fpFr}\end{equation}
where $Q$ and $p$ denote the total and relative momenta respectively,
both ranging in $(-\pi,\pi]$. The eigenvalue of $T$ reflects directly
in the parity of the Fourier transform: $f(-p)=\tau f(p)$, while
the constraint $f_{jj}=0$ becomes $F(0)=\sum_{p}f(p)=0$. When (\ref{fpFr})
is plugged into (\ref{evef}) a common factor $\exp\left({\rm iQ\rho}\right)$
can be simplified from both members and we are left with the simpler
equation:\begin{multline}
(1-\delta_{r,0})\cos\left(\frac{Q}{2}\right)\left[F(r+1)+F(r-1)\right]\\
-\lambda F(r)(\delta_{r,1}+\delta_{r,-1})=\delta E^{\left(2\right)}F(r),\label{eveF}\end{multline}
that appears as a finite-differences Schr\"{o}dinger equation with
a $\delta$-potential at $r=\pm1$. From now on we fix $\tau=1$;
the procedure for $\tau=-1$ is analogous and the conclusions are
exactely the same. In momentum space eq. (\ref{eveF}) becomes:\begin{multline}
4f(p)\cos\left(\frac{Q}{2}\right)\cos p-\frac{2}{L}\left[\lambda\cos p+2\cos(\frac{Q}{2})\right]\\
\times\sum_{p'}f(p'){\rm cos}p'=\delta E^{\left(2\right)}f(p).\label{evefp1}\end{multline}
Were it not for the $\lambda$-term, the spectrum of eigenvalues $\delta E^{\left(2\right)}(Q,p)$
would be identical to that of two non-interacting soliton and anti-soliton
with energy $(2\cos q+2\cos q')=4\cos(Q/2)\cos p$ with $Q=q+q'$
and $p=(q-q')/2$. In the thermodynamic limit $L\rightarrow\infty$
at a given $Q$ these form a continuum $-4\cos(Q/2)\le\delta E_{{\rm cont}}\le4\cos(Q/2)$.
Hence, defining ${\cal E}\equiv\delta E^{\left(2\right)}/4\cos(Q/2)$,
the bound states will be searched in the region $\vert{\cal E}\vert>1$.
Eq. (\ref{evefp1}) can be solved formally as:\begin{equation}
f(p)=\Gamma\frac{C\cos p+1}{\cos p-{\cal E}},\label{fsevefp1}\end{equation}
where $C=\lambda/[2\cos(Q/2)]$ and\begin{equation}
\Gamma\equiv\frac{1}{2\pi}\int_{-\pi}^{\pi}{\rm d}p\: f(p)\cos p\label{defGamma}\end{equation}
Now, re-inserting (\ref{fsevefp1}) into (\ref{defGamma}), self-consistency
demands that either $\Gamma=0$ or:\[
1=\frac{1}{2\pi}\int_{-\pi}^{\pi}{\rm d}p\:\cos p\frac{C\cos p+1}{\cos p-{\cal E}}.\]
Evaluating the integral the energy equation becomes:\[
\sqrt{\frac{{\cal E}^{2}-1}{\vert{\cal E}\vert^{2}}}=\frac{C{\cal E}+1}{C{\cal E}}.\]
With $\lambda>0$ one has a solution either for ${\cal E}>1$ or for
${\cal E}\le\min(-1,-C^{-1})$. Under these conditions the solution
reads ${\cal E}=-(1+C^{2})/2C$. The condition $\mathcal{E}\le-C^{-1}$
should be imposed for $C<1$ but then the requirement would be $C\ge1$.
So we can only accept the case $C>1$ for which $\mathcal{E}<-1$
is always fullfilled. It can be also seen that $\sum_{p}f(p)=0$ is
automatically satisfied if we build $f(p)$ from eq. (\ref{fsevefp1}).
The restriction $C>1$ can be re-expressed as $\cos(Q/2)<\lambda/2$
and defines a characteristic wavenumber $Q^{*}$ above which the bound
state with energy\[
\delta E^{\left(2\right)}(Q)=-\lambda-\frac{4}{\lambda}\cos^{2}\left(\frac{Q}{2}\right),\;\vert Q\vert\ge Q^{*}\]
 emerges from the continuum. This is precisely the result of eq.~(\ref{eq:E2-perturbative}).

\bibliographystyle{/usr/share/texmf/tex/latex/epj/epj}
\bibliography{/home/campos/myart/lambda_D/SPHD}

\end{document}